\documentclass{aa}  

\usepackage{subfigure}
\usepackage{graphicx}
\usepackage{natbib}
\usepackage{txfonts}
\begin{document}
   \title{Extremely high velocity gas from the massive YSOs in IRAS\,17233$-$3606}
   \author{S. Leurini
          \inst{1}          \and C. Codella\inst{2} \and L. A. Zapata\inst{3} \and A. Belloche\inst{3}   \and T. Stanke\inst{1} \and F. Wyrowski\inst{3} \and P. Schilke\inst{3,4}  \and K. M. Menten\inst{3} \and R. G\"usten\inst{3}}

   \offprints{S. Leurini}

   \institute{ESO, Karl-Schwarzschild Strasse 2, 85748 Garching-bei-M\"unchen, Germany\\
  \email{sleurini@eso.org}
	 \and INAF, Osservatorio Astrofisico di Arcetri, Largo E. Fermi 5, 50125 Firenze, Italy 
         \and Max-Planck-Institut f\"ur Radioastronomie, Auf dem H\"ugel 69, 53121 Bonn, Germany 
\and
Physikalisches Institut, Universit\"at zu K\"oln, Z\"ulpicher Str. 77, 50937 K\"oln, Germany
             \\
             }

   \date{\today}

  \abstract
    {Molecular outflows from high-mass young stellar objects provide an excellent way to study the star formation process, and investigate  
if they are scaled-up versions of their low-mass counterparts.}
    {We selected the nearby massive star forming region IRAS\,17233--3606 in order to study the kinematics and physics along the molecular outflow(s) originating from this source.}
    {We observed IRAS\,17233--3606 in CO, a typical tracer of gas associated with molecular outflow, with the Submillimeter Array in the (2--1) transition, and with the APEX telescope in the higher excitation  (6--5) line. Additional infrared H$_2$ observations were performed with the UKIRT telescope. The CO data were analysed using a LVG approach.}
    {Our data resolve the previously detected molecular outflow in at least three different components, one of them with a high collimation factor 
($\sim 4$), and characterised by emission at extremely high velocities ($|v-v_{\rm {LSR}}|> 120$~km~s$^{-1}$). The estimate of the kinematical outflow parameters  are typical of massive YSOs,
and in agreement with the measured  bolometric luminosity of the source. The kinematic ages of the flows are 
in the range $10^2-10^3$ yr, and therefore point to young objects
that still did not reach the main sequence.}
{}

    \keywords{ISM: jets and outflows -- ISM: molecules -- Stars: individual: IRAS 17233-3606 -- Stars: formation}

   \maketitle
%
%

\section{Introduction}\label{intro}

Although massive stars are fundamental for astrophysics, little is
known of their earliest evolutionary phases. The main problem for the
understanding of massive star formation comes from the radiation
pressure that massive stars exert on their surrounding medium once they have reached a mass of around 8~$M_{\rm \sun}$,
that should  prevent further accretion onto the star.  As a
consequence, stars with mass larger than 8~$M_{\rm \sun}$ 
should not exist \citep{1993ApJ...418..414P}, 
in clear disagreement with observations.  Two main theoretical scenarios
have been proposed to solve this problem: (1) accretion as in the
low-mass case through more massive circumstellar disks and/or
sufficiently high accretion rates \citep[e.g.,
][]{2002ApJ...569..846Y} and (2) coalescence of lower-mass stars
belonging to the cluster that harbours the high-mass star(s)
\citep{1998MNRAS.298...93B,2002MNRAS.336..659B}. One of the
fundamental observational tests to distinguish between the different
proposed models is to detect collimated outflows  and accretion
disk from  massive young stellar objects (YSOs), and  to
assess the fraction of still embedded massive YSOs with disks
\citep{2007prpl.conf..197C,2007arXiv0712.0828K}. Since massive
pre-main sequence stars may loose any disk they originally had due to
the radiation field \citep[e.g.,
][]{2000prpl.conf..559N,2003ApJ...598L..39F}, such studies are best
done in earlier evolutionary phases.  However, large distances and a
high degree of multiplicity among early spectral type stars
\citep[e.g.,][]{2001IAUS..200...69P} make observations of these phases of massive YSOs a challenge, especially at (sub)mm
wavelengths where current facilities still lack the necessary
resolution to resolve single massive objects at such distances.

Despite the observational effort in the last decade to investigate the
early evolutionary phases of massive stars, only few sources show
typical phenomena associated with low-mass star formation such as
collimated jets and outflows, circumstellar disks
even at high angular resolution \citep{2005ccsf.conf..105B,2007prpl.conf..245A,2007prpl.conf..197C}. In addition, 
neither collimated jet-like outflows or circumstellar disks have been detected 
in sources more massive than early B- and late O-type YSOs, although  recently \citet{2009ApJ...698.1422Z} detected a 
relatively collimated outflow ($14\degr$) towards W51 IRS2 and suggested the presence of a Keplerian infalling ring around a central object
of at least 60~M$_\odot$. 

It is still not clear whether the lack of detection of collimated outflows and circumstellar disks around O-type YSOs is an observational bias due 
to the poor  angular resolution and sensitivity of the observations or whether it reflects
a real difference in the way massive stars form. 
\citet{2005ccsf.conf..105B} 
suggested an evolutionary scenario in which
the collimation of the outflows decreases with time, 
due to the interaction with the wind from the central object.
Hence,  the need to investigate  high-mass YSOs
in different  evolutionary stages and of different spectral types at high linear resolution is clear.

The region harbouring the prominent far-infrared source
IRAS\,17233$-$3606 (hereafter IRAS\,17233) is one of the best
laboratories in which to attack the problem of massive star formation.
IRAS\,17233 first came to attention
through its very intense H$_2$O, OH, and CH$_3$OH masers \citep[e.g., ][]{1980IAUC.3509....2C,1982ApJ...259..657F,1991ApJ...380L..75M}. The  luminosity of IRAS\,17233 ranges between $
1.4 \times 10^5~L_\odot$ \citep[$d=2.2$~kpc,][based on the IRAS colours]{1993AJ....105.1495H}
and $1.7 \times 10^4~L_\odot$
\citep[$d=0.8$~kpc,][obtained by integrating the spectral energy distribution]{2004A&A...426...97F}.  Previous
studies seem to agree that IRAS 17233 is located at the near kinematic
distance \citep[between 700~pc and
2.2~kpc,][]{2006A&A...460..721M,1989A&A...213..339F} rather than at
the far distance \citep[$\sim 16$~kpc, e.g. ][]{1998AJ....116.1897M}.
This is suggested by the high measured intensities of continuum and line emission
at practically all the observed wavelengths,
which would indicate exceedingly high luminosities if the
source were at the far distance and, persuasively, by the fact that it
is at an angular distance of more than 0.5 degrees below the Galactic
plane. Moreover, the source lies at the edge of an infrared dark cloud (Leurini et al., in prep.).
In the following analysis, we assumed a distance of 1~kpc for IRAS\,17233.

\begin{figure*}
\centering
\includegraphics[angle=-90,width=17cm]{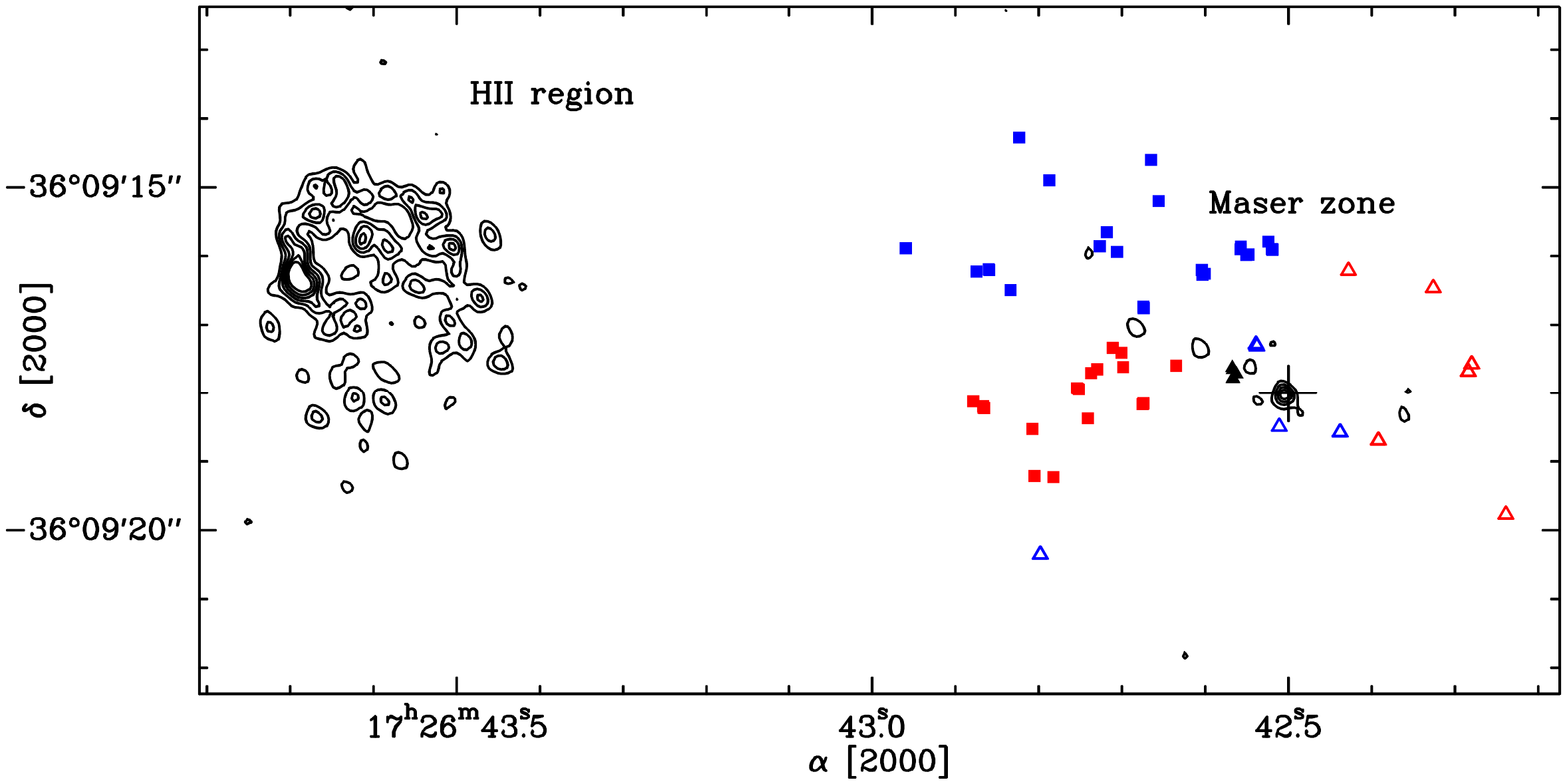}
\caption{Overview of the region: the black contours show the 1.3~cm continuum emission of IRAS\,17233 \citep{2008AJ....136.1455Z},
 from 0.12~Jy~beam$^{-1}$ ($3\sigma$) in step of 
0.24~Jy~beam$^{-1}$. The H{\sc ii} region and the maser zone are labelled.  
Overlaid on the map are the positions of the OH  \citep[blue and red squares,][]{2005ApJS..160..220F},  CH$_3$OH 
\citep[black triangles,][]{1998MNRAS.301..640W}
and H$_2$O \citep[blue and red open triangles,][]{2008AJ....136.1455Z} masers. The cross marks the position of the peak of the SMA 1.3~mm continuum emission, used as reference
for the APEX observations. 
\label{overview}}
\end{figure*}

 Recently, high angular
resolution multi-radio wavelengths observations by
\citet{2008AJ....136.1455Z} resolved a cluster of nine compact radio
sources of different nature in the region. Of these objects, one
corresponds to the  H{\sc ii} region already mapped at
cm-wavelengths by different authors \citep[e.g.,
][]{1993AJ....105.1495H,1998MNRAS.301..640W}, while  four others are
found close to the maser zone (i.e., the area of the region where the H$_2$O, CH$_3$OH and OH masers
are detected, see Fig.~\ref{overview}).  \citet{2008AJ....136.1455Z} also
found that one of these four sources (labelled VLA~2d in their
paper) is at the centre of a bipolar
north-south outflow detected by \citet{2005ApJS..160..220F} in OH
masers.  \citet{2008AJ....136.1455Z} also reported the presence of a ring-like structure
traced by H$_2$O masers, which they interpret as due to
northeast-southwest outflows. \citet{2008A&A...485..167L} reported the
discovery of a bipolar outflow originating from the vicinity of
IRAS\,17233, through low resolution observations performed in CO(3--2)
with the APEX telescope ($\sim 18\arcsec$ resolution). From these data
however, the powering source of the molecular outflow remains
unidentified.  \citet{2008A&A...485..167L} also reported a rich
molecular spectrum from a position very close to the centre of the
outflow, typical of hot molecular cores near massive young stellar objects, that
represent  the stage of massive star formation when the newly formed star is efficiently heating
the surrounding medium up to hundreds of K, but it has not
yet developed an ionised region \citep[e.g., ][]{2000prpl.conf..299K,2003A&A...401..227C}.

In this paper, we present observations of the molecular 
outflow in IRAS\,17233 at 220 and 230~GHz 
in $^{13}$CO(2--1) and CO(2--1) with the Submillimeter Array (SMA), 
and at 690~GHz in CO(6--5)
 with the APEX telescope\footnote{This
publication is based on data acquired with the Atacama Pathfinder
Experiment (APEX). APEX is a collaboration between the
Max-Planck-Institut f\"ur Radioastronomie, the European Southern
Observatory, and the Onsala Space Observatory}. The corresponding continuum data of the SMA observations, as well as 
the other spectral line data, will be presented in a  
forthcoming paper. 
 While the SMA CO(2--1)  data allow us to 
reach a high angular resolution 
(see Table~\ref{obs}) and study the outflow in its detail, the APEX CO(6--5) observations 
trace the warm component of the outflow, and allow a reliable estimate of its excitation conditions. 
Despite the importance of observations of high rotational CO transitions in the study of molecular outflows,
 only a handful of 
sources have been mapped in rotational CO lines with J higher than 3 \citep[e.g., ][]{2002A&A...387..931B,2006A&A...454L..83L,2009A&A...501..633V} because of the low transmission of the atmosphere at these frequencies.

\section{Observations}\label{par-obs}
\subsection{SMA observations}
We observed IRAS\,17233 with the SMA interferometer on April
10, 2007 in the compact configuration with seven
antennas.  The receivers operated in a double-sideband mode with an IF
band of 4-6 GHz so that the upper and lower sideband were separated by
10 GHz. The central frequencies of the upper and lower sideband were
220.4 and 230.4~GHz respectively, allowing to detect the
$^{13}$CO(2--1), C$^{18}$O(2--1) and SO($5_6-4_5$) lines in the lower side
band of the receivers, the CO(2--1) transition in the upper side
band.  The correlator had a bandwidth of $\sim 1.9$~GHz and the
channel spacing was 0.406 MHz, corresponding at this frequency to a
velocity resolution of 0.5~km~s$^{-1}$.  The adopted systemic velocity
was -3.4~km~s$^{-1}$.  The observations were performed under good weather conditions with zenith opacities
$\tau(225~\rm{GHz})$ between 0.1 and 0.14 measured by the National Radio
Astronomy Observatory (NRAO) tipping radiometer operated by the
Caltech Submillimeter Observatory (CSO).

We covered the molecular outflow from
IRAS\,17233 presented by \citet{2008A&A...485..167L} with a
mosaic of two fields centred at
$\alpha_{2000}$=17$^h$26$^m$41$^s$.76,
$\delta_{2000}$=$-$36$^{\circ}$09$'$0$\farcs$5 and
$\alpha_{2000}$=17$^h$26$^m$42$^s$.4,
$\delta_{2000}$=$-$36$^{\circ}$09$'$10$\farcs$5. The projected baselines
ranged between 5 and 88~m. The short baseline cutoff implies that
source structures $\geq 54\arcsec$ are filtered out by the observations.
However, intermediate structures are also affected by missing fluxes.

 Bandpass calibration was done with 3C454.3. We used Callisto for
the flux calibration which is estimated to be accurate within
20\%. Gain  calibration was done via frequent
observations of the quasars 1626-298 and 1713-269. Measured
double-side band system temperatures corrected to the top of the
atmosphere were between 100 and 400\,K, mainly depending on the
elevation of the source.  Details on the observational setup are given in Table~\ref{obs}.

The initial flagging and calibration was done with the IDL superset
MIR originally developed for the Owens Valley Radio Observatory and
adapted for the SMA\footnote{The MIR cookbook by Charlie Qi can be
found at http://cfa-www.harvard.edu/$\sim$cqi/mircook.html.}. The
imaging and data analysis was conducted in MIRIAD
\citep{1995ASPC...77..433S} and
MAPPING\footnote{http://www.iram.fr/IRAMFR/GILDAS}.  Images were
produced using natural weighting. The resulting synthesised beams are listed in Table~\ref{obs}. 
The 
high degree of elongation of the beam is due to the low elevation
of the source from Hawaii. However, the spatial resolution is highest
along the direction perpendicular to the  outflow, making the dataset
well-suited to study its morphology and kinematics.
In order to compare the SMA observations to the APEX data, we resampled the final data to a spectral resolution
of  2~km~s$^{-1}$; the corresponding r.m.s. is 0.05~Jy/beam.

\begin{table}
\centering
\caption{Observational parameters.\label{obs}}
\begin{tabular}{rcccc}
\hline
\multicolumn{1}{c}{Transition} &\multicolumn{1}{c}{Frequency}&\multicolumn{1}{c}{HPBW}&\multicolumn{1}{c}{P.A.}
&\multicolumn{1}{c}{$\Delta v$}\\
& \multicolumn{1}{c}{(GHz)}&
\multicolumn{1}{c}{($\arcsec$)}&\multicolumn{1}{c}{($^\circ$)}&
\multicolumn{1}{c}{(${\rm km~s}^{-1}$)}\\
\hline
$^{13}$CO(2--1)&220.3987&5.43$\times$1.88&30&0.50\\
CO(2--1)&230.5380&5.45$\times$2.08&28&0.50\\
CO(6--5)&691.4731&8.9&--&0.05\\
\hline
\end{tabular}
\end{table}

\subsection{APEX observations}

IRAS\,17233 was observed in the CO(6--5) transition
with the CHAMP$^+$ dual-frequency heterodyne submillimetre array
receiver on APEX. The observations were performed on 2007, October 20 and 22,
 under good weather conditions (with 0.9 and 0.7~mm precipitable
water vapour). The peak of the SMA continuum emission ($\alpha_{\rm
 J2000}=17^h26^m42.455^s$, $\delta_{\rm J2000}=-36^\circ09'18.047\arcsec$) was used as reference position. 
The system temperature was between 4500 and 14\,000~K on October 20th,
between 2800 and 4500~K on October 22nd.
Pointing was established by total power continuum cross-scans
on SgrB2(N), and found to be accurate within $\sim 3\arcsec$.

The beam efficiencies of CHAMP$^+$ were determined via observations of Mars and Jupiter, 
the first at 691~GHz, the second at 661~GHz, and found to be 0.38 and 0.45, respectively. We therefore adopted a beam efficiency of 0.40 to
convert from antenna temperature into main-beam temperature units. The original
spectral resolution of the data (0.05~km~s$^{-1}$) was smoothed to
2~km~s$^{-1}$ for our analysis. The r.m.s. noise in the final data cube 
is not uniform over the channels, because the northern part of the map was observed only on 
 October 20, and because of the different weather conditions during the two days. 
The r.m.s. noise is of the order 1--4~K, larger on the northern part of the map.
The maps were produced with the
XY\_MAP task of CLASS90, which convolves the data 
with a Gaussian of one third  of the beam: the final angular resolution of the data 
is 9$\farcs$4.

\subsection{UKIRT observations}
We obtained near-infrared wide field images using WFCAM on UKIRT
through $JHK$ broad band and a H$_2$ narrow band filter on
May 3rd and 29, 2008. Integration times were 3.75~min for the
broad band filters and 24~min for the narrow band filter. We used
the data as reduced through the CASU pipeline. We used the $K$
broad band image to discriminate between continuum and emission
line features in the H$_2$ narrow band image. We created a continuum
subtracted, pure H$_2$ emission line image by properly registering
the broad and narrow band images and smoothing the narrow band image to
match the poorer seeing of the broad band image. We derived a flux 
scaling factor by comparing the counts of a number of stars in both
filters and subtracted the scaled broad band image from the narrow band
image. The resulting emission line image reveals a number of jet-like
features, of which we will only discuss the counterparts of the CO
outflows here. The other flows from the cluster forming core and
along the filamentary dark cloud will be presented elsewhere (Stanke
et al., in prep.).

\section{Observational results}

\subsection{CO emission}

\begin{figure}
\centering
\includegraphics[angle=-90,width=9cm, bb= 71 51 535 388,clip]{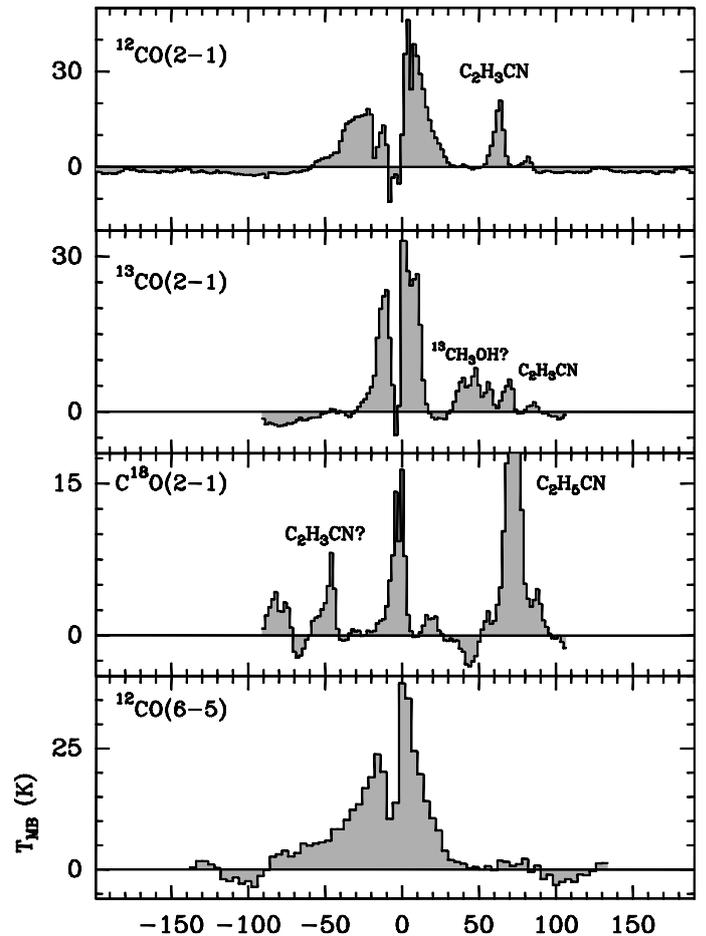}
\caption{Spectra of CO and its isotopologues at the peak of the mm continuum emission; 
the spectral features due to different molecular species are marked. As discussed in
Sect. 3.1,  
the negative platforming, which occurs only in the region where
continuum emission is detected, is probably due to an overestimate of the continuum. In addition, absorption features are probably due to
filtering of large structures. 
 Note also that several hot-core molecular tracers have been detected. Such emissions will be analysed
and discussed in a forthcoming paper focused on the close surroundings of the YSOs.
\label{hc}}
\end{figure}

Our new observations significantly improve the angular resolution of
the previous study \citep{2008A&A...485..167L}, revealing the structure of the molecular outflow
in detail.  The CO transitions are characterised by emission up to
extremely high velocities in the red- and blue-shifted wings.  We
adopt here a systemic velocity of $-3.4$~km~s$^{-1}$ for the source \citep{1996A&AS..115...81B}, which is confirmed by
spectral features in the SMA dataset with peak around
-3.0~km~s$^{-1}$. The blue-shifted emission extends  to $\sim
-200$~km~s$^{-1}$ in the CO(2--1) line, and to $\sim -80$~km~s$^{-1}$
in the CO(6--5) transition.    The
red-shifted lobe extends to velocities up to $120$~km~s$^{-1}$ in
CO(2--1), to $\sim 60$~km~s$^{-1}$ in the (6--5) line.  In Fig.~\ref{hc}, we show the CO
spectra towards the peak of the 1.3~mm continuum emission.

\begin{figure*}
\centering
\includegraphics[angle=-90,width=21cm]{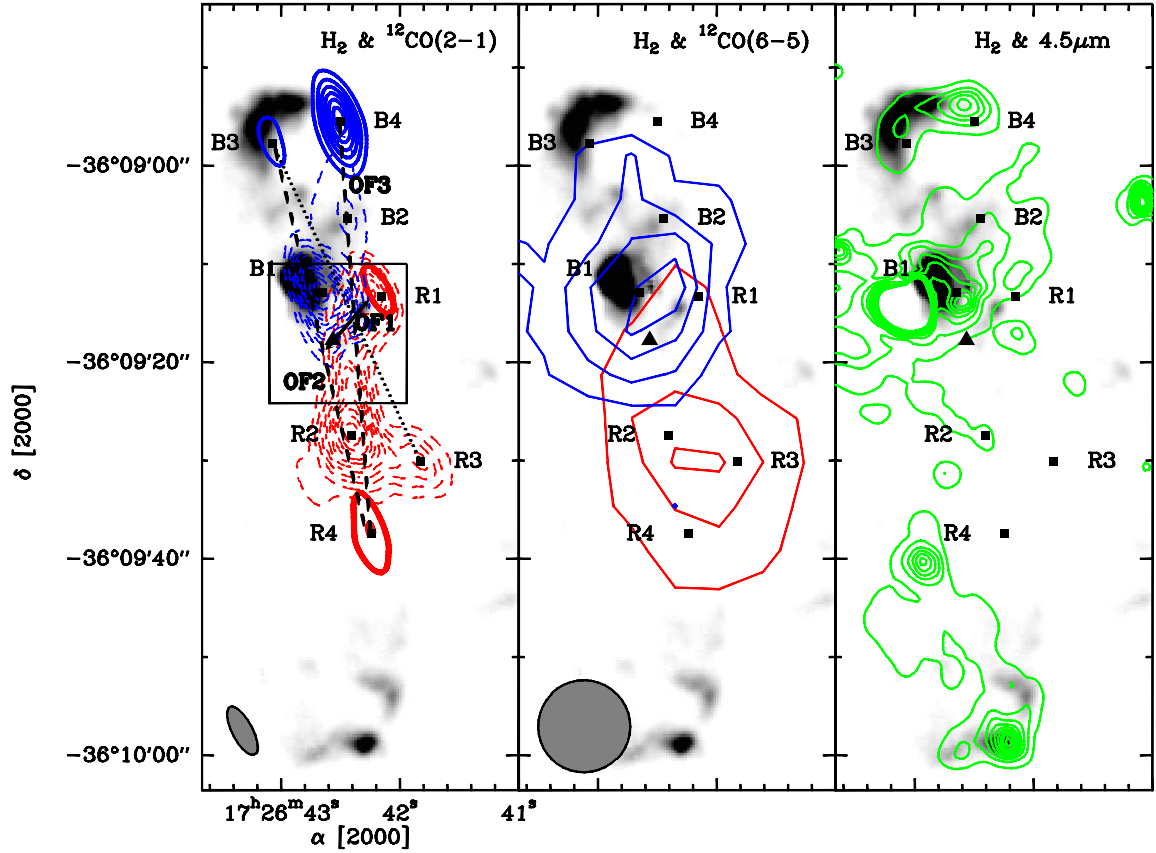}
\caption{Integrated emission of the blue- and red-shifted wings in
the CO(2--1), and (6--5)  lines, and continuum emission at 4.5~$\mu$m overlaid on the H$_2$ emission (grey scale). For the CO(2--1)
transition, the solid contours show the EHV blue-
(v=[-200,-130]~km~s$^{-1}$) and red-shifted emission
(v=[90,120]~km~s$^{-1}$); the dashed contours mark the HV blue-
(v=[-130,-25]~km~s$^{-1}$) and red-shifted emission
(v=[16,50]~km~s$^{-1}$).  For the CO(6--5)  line, the
solid contours show the blue- (v=[-80,-15]~km~s$^{-1}$) and red-shifted emission
(v=[8,60]~km~s$^{-1}$).  All contours start from a $5\sigma$ level and are in step
of $5\sigma$ (see Table~\ref{rms} for the r.m.s noises). The squares mark the positions of
the red-shifted, blue-shifted CO(2--1) peaks, the triangle the CH$_3$OH masers.  
The beam of each observational dataset is
shown in the bottom left corner. In the left panel, the dashed lines and the solid arrow  outline the  possible molecular outflows discussed in \S~\ref{multiple} and labelled IRAS\,17233--OF1, --OF2, --OF3. The dotted line marks the possible connection between R3 and B3. The solid lines in the left panel outline the region mapped in Fig.~\ref{con}.\label{coall}}
\end{figure*}

\begin{figure*}
\centering
\includegraphics[angle=-90,width=19cm]{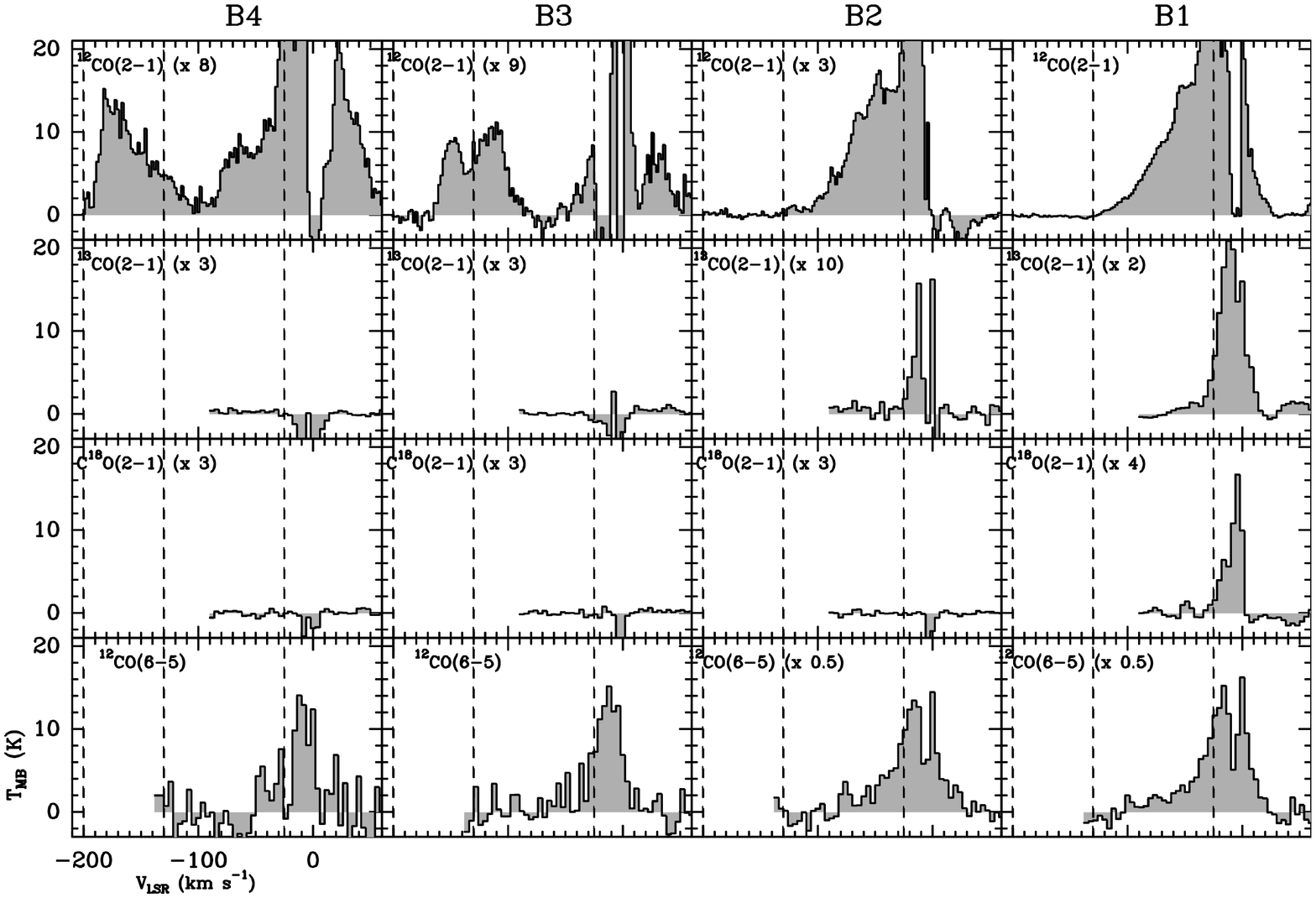}
\caption{Spectra of CO and its isotopologues at the peaks of the CO(2--1) blue-shifted emission. The dashed lines mark the velocities -200, -130 and -25 km~s$^{-1}$ used to define the EHV and HV regimes. Absorption features are probably due to
filtering of large structures. \label{blue}}
\end{figure*}

\begin{figure*}
\centering
\includegraphics[angle=-90,width=18cm]{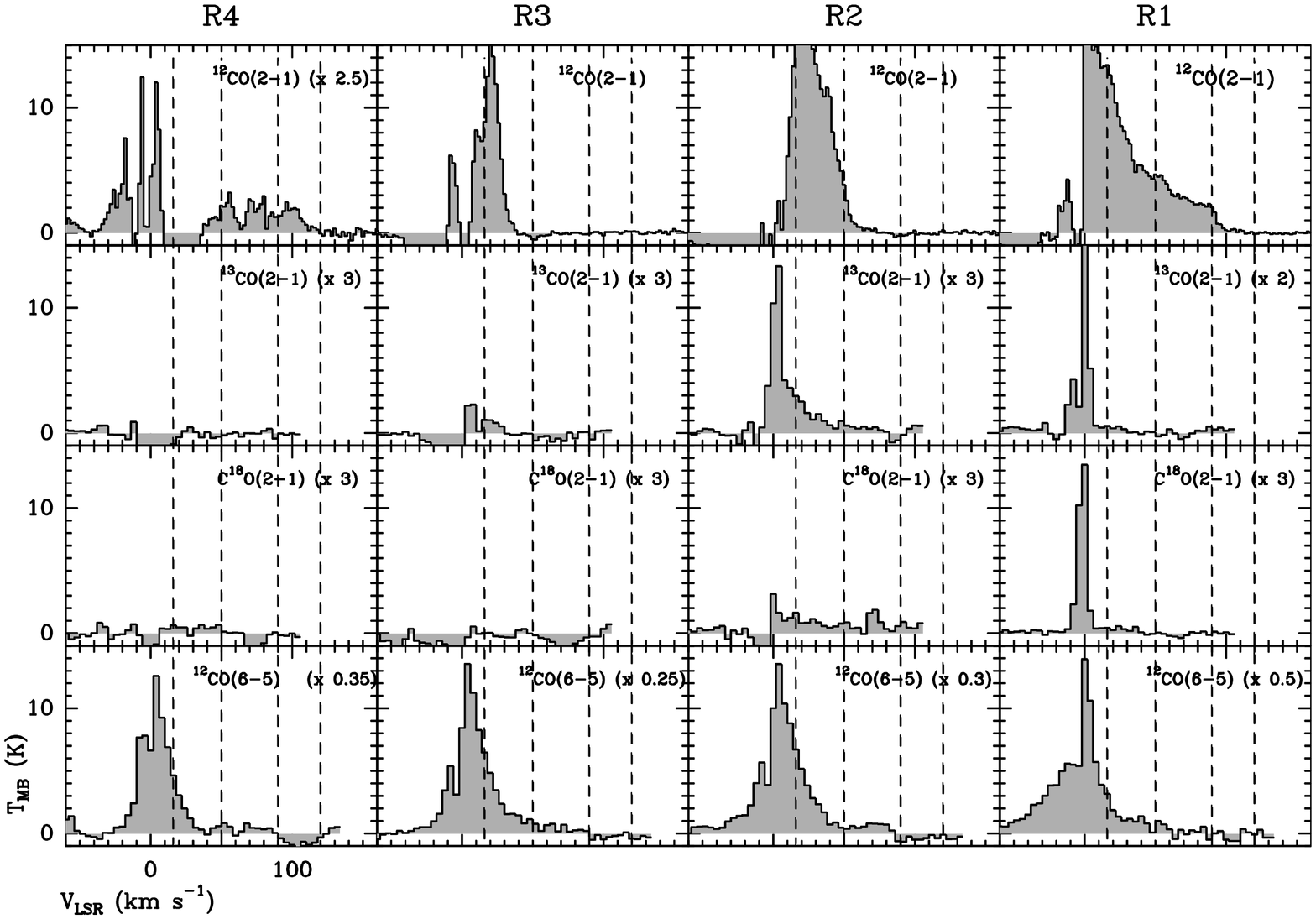}
\caption{Spectra of CO and its isotopologues at the peaks of the CO(2--1) red-shifted emission. The dashed lines mark the velocities 120, 90, 50 and 16 km~s$^{-1}$ used to define the EHV and HV regimes. Absorption features are probably due to
filtering of large structures. \label{red}}
\end{figure*}

Based on the channel maps, we identify two different velocity regimes in the CO(2--1) emission:
the extremely high velocity regime (EHV: blue: $-200<v_{\rm
LSR}<-130$~km~s$^{-1}$; red: $90<v_{\rm LSR}<120$~km~s$^{-1}$), and
the high velocity regime (HV: blue: $-130<v_{\rm LSR}<-25$~km~s$^{-1}$;
red: $16<v_{\rm LSR}<50$~km~s$^{-1}$). The integrated intensity in the
two velocity intervals is presented in the left panel of
Fig.~\ref{coall}. For the high velocity red-shifted emission, we did
not integrate over the velocities $(50,90)$~km~s$^{-1}$ because of
contamination from other molecular species at the peak of the mm
continuum emission (see Fig.~\ref{hc}). We stress here that the
molecular spectrum towards the peak of the continuum emission is
extremely rich and typical of a hot-core, with very few channels free
of line emission. For this reason, we cannot exclude contamination
from other molecular species, but this problem should affect our maps
only at the centre of the mm continuum emission.  The negative
platforming seen in the spectra towards this position (Fig.~\ref{hc})
is probably a result of at least two effects: first, it is not unlikely
that we overestimate the contribution of the continuum, since at the
position of the hot core the density of lines is extremely high and it
is difficult to define channels free of line emission. Since the
negative platforming affects only the area of the map where the
continuum emission is detected (see Fig.~\ref{blue} and \ref{red} for
comparison of spectra at other positions), we believe this to be the
most likely reason for it. However, higher energy transition lines
(e.g., CO(6--5) but also CO(3--2), \citealt{2008A&A...485..167L}) have
wing emission at very high velocities, and therefore it is possible
that large structure emission is filtered out by the SMA and results
in negative features.

\begin{figure*}
\centering
\includegraphics[angle=-90,width=21cm]{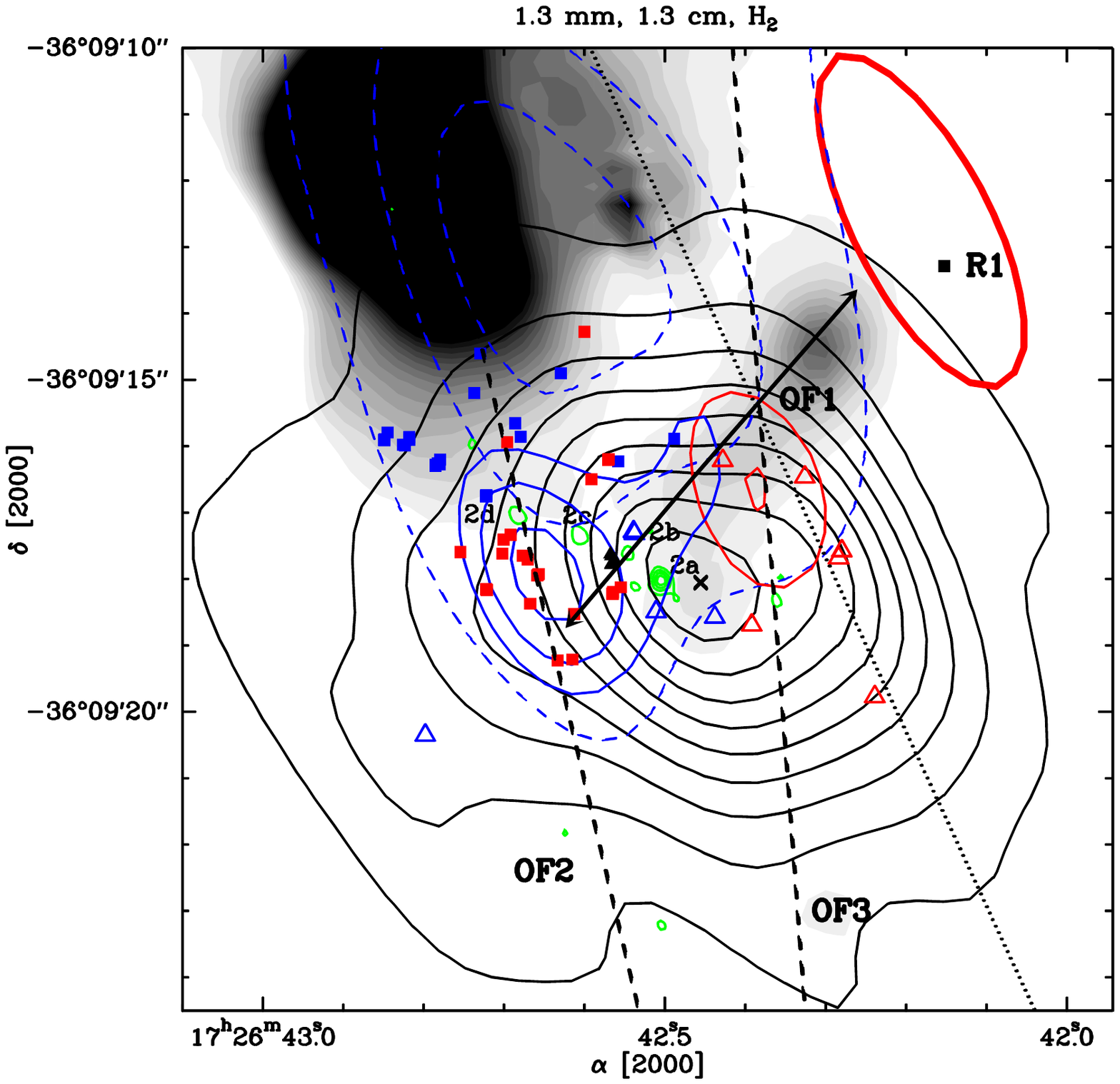}
\caption{Continuum emission at 1.3~mm observed with the SMA (Leurini et al., in prep.) from the maser zone in
  IRAS\,17233 (solid black contours from 0.12~Jy~beam$^{-1}$
  ($3\sigma$) in step of 0.24~Jy~beam$^{-1}$).  Overlaid on the map
  are the positions of the OH \citep[blue and red 
    squares,][]{2005ApJS..160..220F}, of the CH$_3$OH \citep[black
    triangles,][]{1998MNRAS.301..640W} and of the water \citep[blue
    and red open triangles,][]{2008AJ....136.1455Z} masers. The green
  contour levels show the continuum emission at 1.3~cm \citep[from
    1.4~mJy~beam$^{-1}$ ($3\sigma$) in step of
    1.4~mJy~beam$^{-1}$,][]{2008AJ....136.1455Z}, while the grey scale
  map is the H$_2$ emission. The dashed blue contour levels represent the CO(2--1) blue-shifted emission at high velocities 
(from $5\sigma$ in step of 20$\sigma$, see Table~\ref{rms}), the solid blue and red contours the SO($5_6-4_5$) blue- and red-shifted emission 
(blue: from 80 to 104~Jy~beam$^{-1}$ in step of 10~Jy~beam$^{-1}$; red: 10 and 12~Jy~beam$^{-1}$).
The black square marks the R1 position,
  detected in the EHV red-shifted CO(2--1)
  emission (thik solid red contour, as in
  Fig.~\ref{coall}). The cross marks the position of the peak position of the 1.3~mm continuum emission. The dashed and dotted lines and the arrow are as in Fig.~\ref{coall}.  OF1, OF2, and OF3 label the three identified outflows. \label{con}}
\end{figure*}

The SMA CO maps reveal a clumpy structure with  well separated blue- and red-shifted emission, and
an overall structure well aligned along the N-S direction and centred close
to the peak of the mm continuum emission ($\alpha_{\rm
J2000}=17^h26^m42.455^s$, $\delta_{\rm
J2000}=-36^\circ09'18\farcs047$, see Fig.~\ref{con}). This peak is associated with the maser zone
 and not with the H{\sc ii} region reported by
several authors (see Fig.~\ref{overview}).  The red-shifted emission shows four peaks: R1 associated with HV and EHV
emission,
R2 and R3 only with HV gas, and R4 with EHV emission only.  Similarly, the blue-shifted
emission peaks at four positions: B1 and B2, associated only with HV gas, and B3 and B4 associated with EHV emission. 
The spectra at the eight peaks of the CO(2--1) emission are shown in 
Figs.~\ref{blue}-\ref{red}.

Figure~\ref{coall} also shows the integrated emission of the blue- and
red-shifted wings of the CO(6--5)  line. Despite the
different resolutions, the two maps are very similar. Emission is
detected at all positions detected in CO(2--1), except on the
high-velocity B4.  

\begin{table}
\centering
\caption{Noise level in the integrated intensity maps presented in Fig.~\ref{coall} and \ref{others}.\label{rms}}
\begin{tabular}{lcr}
\hline
\multicolumn{2}{c}{Transition} &\multicolumn{1}{c}{rms}\\
&&\multicolumn{1}{c}{[Jy~beam$^{-1}$]}\\
\hline
CO(2--1)& red-shift., HV  &0.5\\
CO(2--1)& red-shift., EHV &1.3 \\
CO(2--1)& blue-shift., HV &8.0 \\
CO(2--1)& blue-shift., EHV &1.3 \\
CO(6--5)& red-shift.&1035  \\
CO(6--5)& blue-shift.&863  \\
$^{13}$CO(2--1)&red-shift.&0.3\\
$^{13}$CO(2--1)&blue-shift.&1.0\\
C$^{18}$O(2--1)&red-shift.&0.4\\
C$^{18}$O(2--1)&blue-shift.&0.7\\
SO($5_6-4_5$)&red-shift.&0.4\\
SO($5_6-4_5$)&blue-shift.&2.0\\
\hline
\end{tabular}
\end{table}

To roughly quantify the degree of collimation of the whole outflow structure
as traced by CO(2-1), we divided the length of the
outflow by its width, which results in a collimation factor $f_c\sim
4$. In doing so, we derived the length (B4--R4) and width
(B4--B3) of the flow from the $5\sigma$ level contour of the EHV
integrated intensity, and assumed that this comes from only one
molecular outflow (see  \S~\ref{multiple} for a discussion on the
multiplicity of flows in the region). Similarly, the opening angle of
the outflow can be estimated from the length of the deconvolved  major axis $a$ and
the length of the deconvolved  minor axis $b$, $\delta=2\times$arctan $(b/a)$. The derived value is $\delta\sim
11\degr$. Alternatively, if the H$_2$ emission in the south of R4 is associated with the same
outflow, the collimation factor would be $\sim 6$ and the opening angle $8\degr$. Assuming a
distance of 1~kpc, the projected extension of the outflow on sky is 0.2--0.3~pc.

\subsection{CO outflow vs. H$_2$ emission}

Figure \ref{coall} compares the red- and blue-shifted CO emission with the H$_2$ at 2.12~$\mu$m map and the 
 4.5~$\mu$m continuum emission from the Spitzer Space Observatory \citep{2003PASP..115..953B}.
This emission is often attributed to H$_2$ lines tracing shocked gas arising from outflow activity \citep{2004ApJS..154..352N,2009ApJ...695L.120Y}, 
although contamination from rovibrational lines of  CO may also happen in hot gas  \citep{2004ApJS..154..333M}. 
H$_2$ emission at 2.12~$\mu$m is found in correspondence to all the blue-shifted clumps. Emission 
at 4.5~$\mu$m 
is detected towards both B3 and B4.
Interestingly, the EHV B3 and B4 clumps are located at the edges of the IR bow-structure traced by H$_2$ and by the
4.5~$\mu$m emission, 
and 
could trace material recently shocked.  

Probably due to extinction, only R1 among the red-shifted clumps
shows an H$_2$ counterpart (see discussion in paragraph \S~\ref{multiple}). Figure~\ref{coall} shows also H$_2$ emission
$\sim$ 40$\arcsec$ south of the YSOs, but it is not clear whether it
is tracing emission related to the red-shifted IRAS\,17233 outflow.
The EHV R1 clump 
lies ahead of  an independent, short H$_2$ jet (see Fig.~\ref{con}): also in this
case it seems we are observing material accelerated along the jet traced by H$_2$
(see  \S~\ref{multiple}).

\subsection{Emission from other molecular species}

\begin{figure*}
\centering
\includegraphics[angle=-90,width=16cm]{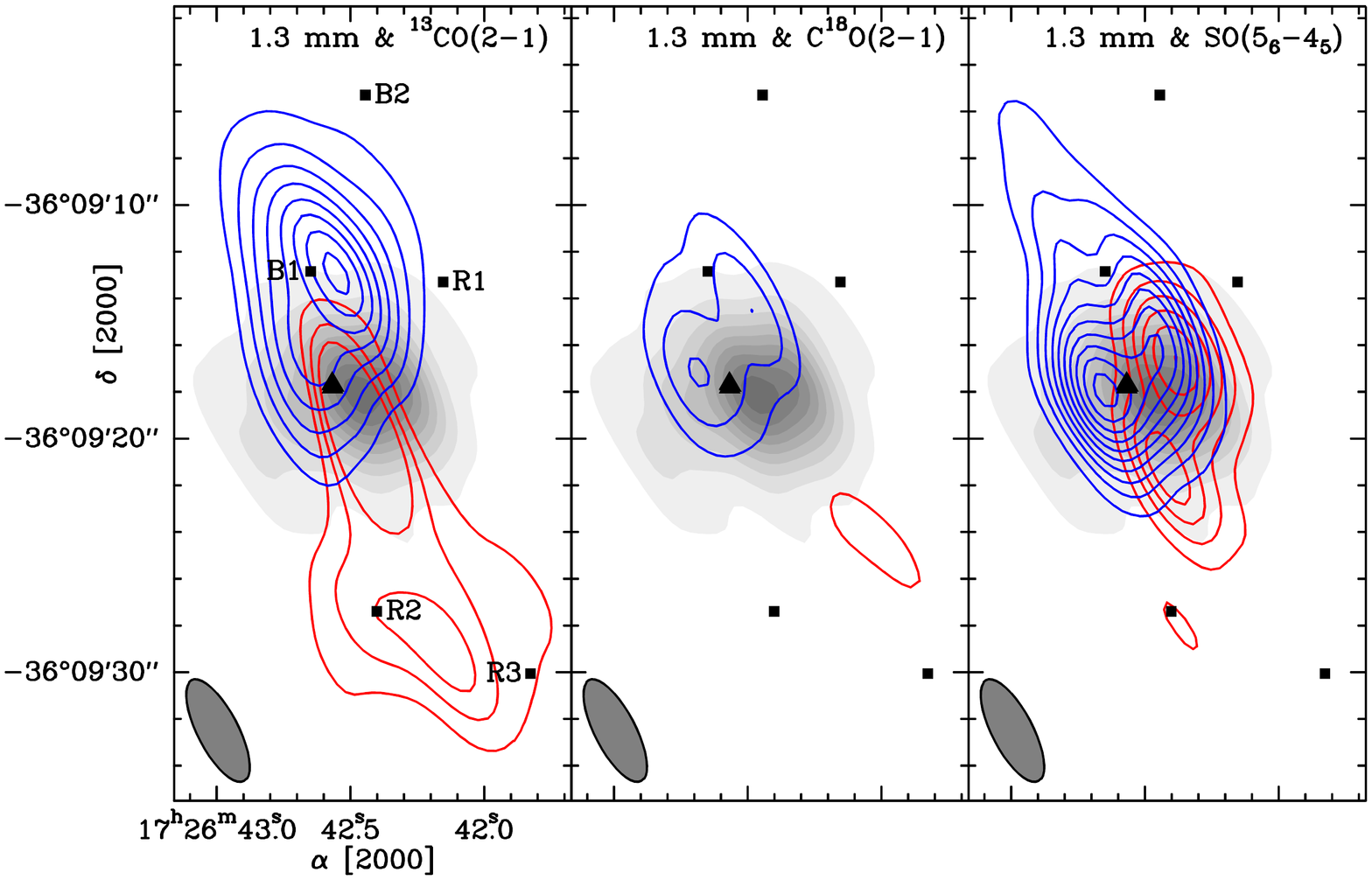}
\caption{Maps  of the integrated blue- and red-shifted emission in
the $^{13}$CO(2--1), C$^{18}$O(2--1) and SO($5_6-4_5$)  transitions. The solid contours show the
blue-shifted emission  in the velocity ranges v=$[-55,-20]$~km~s$^{-1}$ for the $^{13}$CO(2--1) line, 
v=$[-20,-11]$~km~s$^{-1}$ for C$^{18}$O(2--1) and v=$[-40,-11]$~km~s$^{-1}$ for SO($5_6-4_5$), and the
red-shifted emission  in the velocity range v=$[15,20]$~km~s$^{-1}$ for the $^{13}$CO(2--1) line, 
v=$[5,10]$~km~s$^{-1}$ for C$^{18}$O(2--1) and v=$[15,25]$~km~s$^{-1}$ for SO($5_6-4_5$).
The contours start from a $5\sigma$ level and are in step of $5\sigma$. The grey scale map is the continuum emission
at 1.3~mm (from 0.12~Jy~beam$^{-1}$
  ($3\sigma$) in step of 0.24~Jy~beam$^{-1}$).
The symbols are as in Fig.~\ref{coall}. 
The  synthesised beam  is shown in the bottom left corner.\label{others}}
\end{figure*}

Together with the main isotopologue of CO, also the $^{13}$CO,
C$^{18}$O(2--1), and  SO($5_6-4_5$) transitions show
non-Gaussian wings in the spectrum taken at the position of the mm
continuum emission (Fig.~\ref{hc}).   Figure~\ref{others} shows the
integrated intensity of the three transitions in the red- and
blue-shifted wings.  For the red-shifted emission $^{13}$CO emission, we only used the range of velocities $(15,20)$~km~s$^{-1}$ 
because of contamination from other molecular species.

The emission traced by the three molecular
species is more compact than that traced by $^{12}$CO, as expected. For the red-shifted
emission, the $^{13}$CO emission extends towards the features R2 and
R3, while the SO($5_6-4_5$) transitions has a $5\sigma$ detection on
R2 and extends towards the R1 position, where EHV 
red-shifted CO(2--1) and H$_2$ are detected. For the blue-shifted
emission, all transitions are detected on B1, although none of them
peaks at this position.  All three
transitions show  two spatially separated blue-shifted emission peaks; the same
behaviour is seen in the CO(2--1) line at velocities
between --40 and --20 km~s$^{-1}$, similar to the range of velocities used to obtain  
the blue-shifted integrated maps of $^{13}$CO and SO. However, the position of the two
peaks is not the same for the four species. The east blue-shifted peak is probably associated with an outflow
traced by a H$_2$ jet and by the H$_2$O masers (see discussion in the following paragraph); the west peak is probably related to one of the other 
two outflows of the region.

\section{Multiple molecular outflows from IRAS\,17233}\label{multiple}

A zoom in the central region of the molecular outflow reveals a
complex picture. Figure~\ref{con} shows the continuum emission at
1.3~mm towards the inner region of the outflow detected in CO;
overlaid on the map are the 1.3~cm continuum emission from
\citet{2008AJ....136.1455Z}, the positions of the OH, H$_2$O and
CH$_3$OH maser spots
\citep{2005ApJS..160..220F,2008AJ....136.1455Z,1998MNRAS.301..640W}, the CO(2--1) HV blue-shifted emission, the SO blue- and red-shifted emission, 
and the H$_2$ emission. 

Four compact sources are detected in cm
continuum emission within $\sim 2\arcsec$ from the peak of the mm continuum
emission \citep[and labelled VLA~2a, 2b, 2c and 2d by][]{2008AJ....136.1455Z}. 
The brightest source, VLA~2a, is at  $0\farcs7$ from the mm
continuum peak, suggesting that the two sources probably
coincide.
The cm VLA~2a emission reveals two components,
one dominated by a classical H{\sc ii} region, the other likely due to
dust emission.  
On the other hand, VLA~2b shows a negative spectral index, possibly
associated with synchrotron emission. 
Finally, VLA~2c and 2d have  a spectral energy distribution  
compatible with a hyper-compact
H{\sc ii} region (HCH{\sc ii}) or with a massive dusty core or disk. 

Other information on the sources comes from the maser emission. 
The CH$_3$OH masers (black triangles in Fig.~\ref{con}), so far detected only towards massive stars
\citep[e.g., ][]{2008A&A...485..729X},   
seem to indicate their association with VLA~2b, although  with an uncertainty of $\sim 1\arcsec$ in the absolute position. 
A bipolar outflow is detected in OH, with blue-shifted
emission towards the north and red-shifted emission towards the south
as for the CO emission. OH masers (filled squares) are
observed from molecular material surrounding ultra-compact H{\sc ii}
regions, and from protostellar outflows associated or not with cm
continuum \citep[e.g., ][]{2003ApJ...593..925A}. In the case of OH masers
tracing molecular outflows, the non detection of an ultra-compact
H{\sc ii} region can be explained with a very early evolutionary
stage of the source, or with an intermediate mass object which will
never produce a detectable ultra-compact H{\sc ii} region. 
In our case, VLA~2d could be the driving source of this outflow. 
Finally, we have a pattern formed by red- and blue-shifted H$_2$O
masers (empty triangles), which are well known tracers of molecular jets/outflows,
 spread around the 1.3 mm continuum peak.

Given the current resolution of
our CO observations and the multiplicity of the high-velocity clumps, we cannot
derive the exact number of
molecular outflows in the region, in particular those
extended along the N-S direction.
However, using the  high angular 
resolution  H$_2$ and maser emission data,  we can try to draw a scenario where 
at least three molecular outflows originate from
the maser zone in IRAS\,17233: 

\begin{enumerate}

\item
the first one (IRAS\,17233--OF1, indicated by a solid arrow
in Figs.~\ref{coall} and \ref{con}) is well traced by a H$_2$ jet  which seems to arise from the 
1.3~mm continuum emission peak: the driving source
could be either VLA~2a or VLA~2b. In addition, the H$_2$O spots are
well in agreement with a molecular outflow directed along the
NW(red)-SE(blue) direction.  The present $^{12}$CO maps confirm this
possibility, with high-velocity blue emission located SE of the 1.3~mm
peak (Fig.~\ref{con}), and the R1 red clump which is located ahead the
H$_2$ NW jet. This picture is consistent with the scenario depicted by SO, a well-known
molecule whose abundance is enhanced in molecular outflows
\citep[e.g.,][]{1997ApJ...487L..93B,2005MNRAS.361..244C}: blue-shifted
emission east of the 1.3~mm peak, and red-shifted emission in the
NW (Fig.~\ref{con}).
Blue-shifted emission is also detected in
$^{13}$CO and C$^{18}$O (Fig.~\ref{others}) in the SE of the 1.3~mm
peak.  Thus, the R1 clump seems to trace EHV gas associated with the
mass loss traced by H$_2$.  The scenario for the SE blue lobe is less
clear: we have high velocity blue emission but no EHV clump and,
surprisingly, there is no H$_2$ counterpart, when usually the blue
H$_2$ emission is brighter than the red one due to extinction.  A possible explanation would be that 
the blue-shifted jet is leaving the high density clump hosting the YSOs, moving in  
a gas  where the density is so low that H$_2$ is not excited.
A more realistic solution is that the driving source of the H$_2$ NW 
jet (VLA~2a or VLA~2b) had formed inside the 1.3~mmm core, but close
to its far edge; in this case the red-shifted jet would travel in a less dense ambient
medium, while the blue-shifted jet would travel deeply into the molecular
cloud with a consequently higher extinction.

\item
a second outflow can be found along the direction traced by the
B3-B1-R2-R4 clumps (IRAS\,17233--OF2, dashed line in Figs.~\ref{coall} and
~\ref{con}). Note that this direction is well aligned with both the OH
bipolar outflow and definitely supports VLA~2d as driving source;

\item
the third one can be found along the path traced by the 
B4-B2-R2-R4 clumps (IRAS\,17233--OF3). In this case, the corresponding dashed line
of Figs.~\ref{coall} and \ref{con} passes through the H$_2$O masers 
and close to the VLA~2a/1.3mm source,
which could then host the driving source.

\end{enumerate}

This picture has to be considered as a first incomplete step to prepare
higher angular resolution observations, in CO or in a classical tracer of the
hot jet component as SiO, which are needed for a better understanding
of the number of outflows in the region, and for the identification of
the powering sources. For instance, there is no clear solution
accounting for the R3 clump.  A possible blue counterpart of R3 could
be found in B3 and the driving source should be consequently in the
1.3~mm object. Similarly, R2 and R4 are not clearly assigned to any given outflows, but in our current picture 
they could either belong  to  IRAS\,17233--OF2 or to IRAS\,17233--OF3.

In conclusion, although higher angular resolution data are needed for
a better understanding of the number of outflows in the region, and
for the identification of the powering sources, on the basis of the
existing data we can safely conclude that at least three molecular
outflows are found in the maser zone in IRAS\,17233. Given their
association with CH$_3$OH and OH maser emission, and given the
presence of at least four compact sources of cm continuum emission, we
conclude that the molecular outflows originate from intermediate  or
high-mass YSOs.

\section{Derived outflow parameters}

Global properties of outflows can be derived,
under given assumptions, from the high-velocity CO emission. 
We will use the $^{13}$CO(2--1), $^{12}$CO(2--1) and
$^{12}$CO(6--5) data to infer the optical depth and the
excitation temperature. However, there is 
a major caveat in our analysis: interferometric data are affected by
missing flux from extended structures. Moreover, optically thick lines
have a more uniform distribution than optically thin transitions, and
may be more sensitive to interferometric spatial filtering. Thus, in the following discussion we may {\it
(1)} overestimate the optical depth of the $^{13}$CO(2--1) line; {\it
(2)} overestimate the ratio between the $^{12}$CO(6--5) and the
$^{12}$CO(2--1) lines, and therefore {\it (3)} overestimate the excitation temperature of the gas.  In both cases, the problem comes
from the
$^{12}$CO(2--1) transition, whose  observed flux may be a lower limit to its true
value. Since no single dish $^{12}$CO(2--1) observations of IRAS\,17233
 are available, we cannot further investigate this possibility. In section \S~\ref{tau}, we will derive the optical depth of the $^{13}$CO(2--1)
and $^{12}$CO(2--1) lines; in \S~\ref{tex}, we will investigate the excitation conditions of the gas along the outflow. Finally, in \S~\ref{para}
we will derive the properties of the outflow making use of the opacities derived in \S~\ref{tau} and for a range of temperatures compatible
with the results of the statistical-equilibrium calculations performed in section \S~\ref{tex}.

\subsection{Optical depth}\label{tau}

\begin{figure}
\centering
\includegraphics[angle=-90,width=15cm]{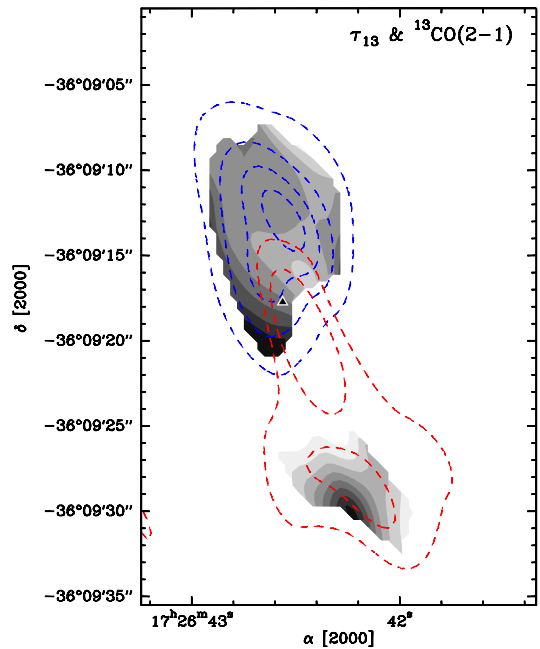}
\caption{Overlay of the opacity of the $^{13}$CO wings ($\tau_{\rm 13}$; grey scale, levels are from 0.07 (light grey) to 0.12 (dark grey) 
in step of 0.01)
as averaged in the --55,--20 km s$^{-1}$ (red) and +15,+25 km s$^{-1}$ (blue) ranges,
and the $^{13}$CO integrated blue- and red-shifted outflow (see Fig.~\ref{others}: for the sake of clarity only
selected contour levels are drawn). The triangle marks the position of the
CH$_3$OH methanol masers (see Fig.~\ref{con}). \label{tau13}}
\end{figure}

By comparing the intensity of the $^{13}$CO and $^{12}$CO(2--1)
intensities in the wings, it is possible to estimate the optical
depths of both lines, assuming a [$^{12}$CO/$^{13}$CO] relative
abundance of 77 \citep{1994ARA&A..32..191W}. We estimated the optical
depth of the $^{13}$CO(2--1) line, $\tau_{\rm 13}$, under the
assumptions that the $^{13}$CO(2--1) and $^{12}$CO(2--1) lines have
the same excitation temperature, and that the $^{12}$CO(2--1)
transition is optically thick.  Since (i) the S/N ratio in the
$^{13}$CO(2--1) wings is sufficiently high only in the relatively low
velocity outflow range, and (ii) self-absorption affects the
$^{12}$CO(2--1) line at velocities close to the ambient one, we
derived the optical depths in the --55,--20 km s$^{-1}$ (red) and
+15,+25 km s$^{-1}$ (blue) ranges.  Figure \ref{tau13} shows the
distribution of the optical depth of the $^{13}$CO(2--1) line averaged
over these velocity ranges: the emission in the $^{13}$CO(2--1) line
results to be optically thin, with $\tau_{\rm 13}$ $\le$ 0.1-0.3. The
$^{12}$CO(2--1) optical depth, $\tau_{\rm 12}$, was derived by simply
scaling the $\tau_{\rm 13}$ value by the assumed [$^{12}$CO/$^{13}$CO]
ratio, thus resulting in an optical depth for the $^{12}$CO(2--1)
line, $\tau_{\rm 12}$, between 10 and 30.

\subsection{LVG analysis}\label{tex}

\begin{figure}
\centering
\includegraphics[angle=-90,width=15cm]{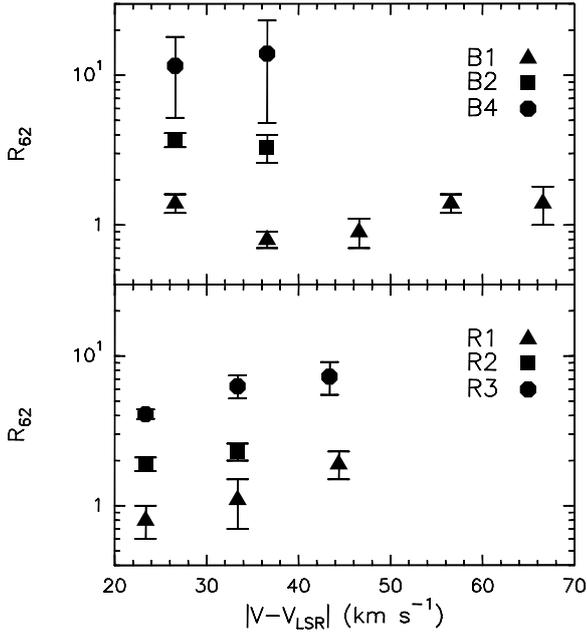}
\caption{Distribution with velocity with respect to the systemic one ($V_{\rm LSR}$) of the ratio
$R_{\rm 62}$ between the brightness temperatures of the $J$ = 6--5 and 2--1 lines as observed
towards the blue- (upper panel) and red-shifted (lower panel) outflow lobes.  \label{wings}}
\end{figure}

\begin{figure}
\centering
\includegraphics[angle=-90,width=11cm]{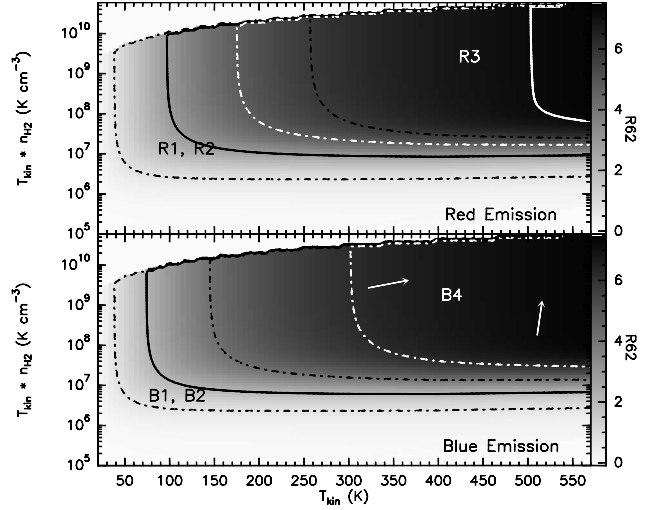}
\caption{Results of statistical equilibrium calculations for $^{12}$CO. The ratio $R_{\rm 62}$ 
(grey scale) between the brightness temperature of the $J$ = 6--5 and 2--1 lines,
as a function of the $T_{\rm kin}$ $\times$ $n_{\rm H_2}$ product, which can be used as a measure
of the gas pressure, and $T_{\rm kin}$. 
We have assumed the LVG optical depth parameter $N_{\rm CO}$ $\Delta$$v$$^{-1}$
= 10$^{16}$ cm$^{-2}$ (km s$^{-1}$)$^{-1}$. Solid lines mark $R_{\rm 62}$
as measured towards the red- (upper panel) and blue-shifted (lower panel) outflow lobes. 
Dot-dashed lines are for the $R_{\rm 62}$ uncertainties. Note that the $R_{\rm 62}$ value towards
B4 is out of the sampled plane ($\sim$ 13): white arrows are drawn to clarify the portion of the plane allowed
for B4.\label{lvg}}
\end{figure}

In order to constrain the physical conditions associated with the CO
outflow(s) by means of statistical-equilibrium calculations, we
compared the $^{12}$CO $J$~=~6--5 and 2--1 lines ($E_{\rm u}$~=~116.2 and 16.6 K, respectively). 
 For a proper comparison, the $^{12}$CO(2--1) image,
originally obtained with a 5$\farcs$5 $\times$ 2$\farcs$1 resolution,
was reconstructed at the lower resolution of the $^{12}$CO(6--5) map
(HPBW = 9$\farcs$4).  We then measured the ratio $R_{\rm 62}$
between the brightness temperature of the $J$~=~6--5 and 2--1 lines.
Figure \ref{wings} shows the distribution of $R_{\rm 62}$ with
velocity with respect to the systemic one ($V_{\rm LSR}$) as observed
towards the blue- (upper panel) and red-shifted (lower panel) outflow
lobes.  Interestingly, Fig. \ref{wings} shows that, for each blue- and
red-shifted lobe the higher the distance of the clump from the
driving source(s) is, the higher is the $R_{\rm 62}$ ratio, thus
indicating higher excitation conditions.  For instance, the $R_{\rm
62}$ ratio calculated for the B4 clumps is very high: $\sim$ 10.  However, the CO emission is associated with
 multiple outflows, and among the CO peaks only B2 and B4 clearly appear to be related to the same flow. Therefore, 
the trend seen in the $R_{\rm 62}$ ratio may not be physically relevant for the other positions. Higher angular resolution observations are needed 
to get a clearer picture of the region. In
addition, Fig. \ref{wings} suggests also, at least for the red clumps,
an increase of the excitation at the highest velocities, probably
closely associated with recently shocked gas.

In order to constrain the excitation conditions, 
we used a Large Velocity Gradient (LVG) code with collisional rates from \citet{2005A&A...432..369S}. These
include extrapolation to energy levels up to J=40 and collisional temperatures up to 2000 K, based on the original
datasets calculated  by \citet{2001JPhB...34.2731F} and \citet{2006A&A...446..367W} for 
temperatures in the range from 5 to 400 K and energy levels up to J=29 and J=20 for collisions with para-H2 and ortho-H2, respectively.

We assumed a specific column density $N_{\rm CO}$ $\Delta$$v$$^{-1}$
ranging from 10$^{14}$ to 10$^{19}$ cm$^{-2}$ (km s$^{-1}$)$^{-1}$. 
Figure \ref{lvg} shows our typical model results, illustrated graphically
with the ratio $R_{\rm 62}$ (grey scale) as a function of the thermal pressure 
$(T_{\rm kin}$ $\times$ $n_{\rm H_2})$ and the kinematic temperature $(T_{\rm kin})$. Figure \ref{lvg}
shows the solutions found for $N_{\rm CO}$ $\Delta$$v$$^{-1}$ 
= 10$^{16}$ cm$^{-2}$ (km s$^{-1}$)$^{-1}$: however, this should
be considered as an upper limit since also solutions down to
 10$^{14}$ cm$^{-2}$ (km s$^{-1}$)$^{-1}$ are acceptable. 
Solid lines mark the measured values of $R_{\rm 62}$ towards the red- 
(upper panel) and blue-shifted (lower panel) outflow lobes.
For the sake of clarity we plotted the values averaged over the whole velocity ranges. 
In addition, we further averaged the measurements of the clumps associated with the lower
excitation conditions, showing similar values 
(red: $R_{\rm 62}$(R1,R2) = $(R_{\rm 62}$(R1)+$R_{\rm 62}$(R2))/2; 
blue:  $R_{\rm 62}$(B1,B2) = $(R_{\rm 62}$(B1)+$R_{\rm 62}$(B2))/2, where  $R_{\rm 62}$ at 
each position is the average over the values 
for different velocities). 
Dot-dashed lines are for the $R_{\rm 62}$ uncertainties, defined as the minimum and maximum values that 
$R_{\rm 62}$(R1,R2) reaches. 

Figure \ref{lvg} clearly shows that  
we cannot obtain severe constraints on the physical conditions of the gas with only two transitions.
However, it  allows to quantify the different excitation conditions
of the outflow clumps. Regarding the red lobe, the R1 and R2 clumps
have $T_{\rm kin}$ $\ge$ 50 K and a product 
$T_{\rm kin}$ $\times$ $n_{\rm H_2}$, which can be considered a measure of the gas pressure,
$\ge$ 10$^6$ K cm$^{-3}$.  
A similar scenario can be found in the blue lobes  B1 and B2.
On the other hand, the $R_{\rm 62}$ value towards the extreme high velocity position
B4 is out of the plane sampled by our models ($R_{\rm 62}\sim$ 10). The red-shifted extreme high velocity position
(R3, $R_{\rm 62}\sim$ 7) is at the limit of the plane.
Such high values of $R_{\rm 62}$ suggest very high temperatures ($\ge 150$~K for R3 and $\ge 300$~K for B4) 
for the gas at those positions;
however, these values
must be taken as upper limits since the flux of the CO(2--1) line could be underestimated due to filtering of large
structures, as stated in Sect. 5.

\subsection{Outflow energetics and kinematics}\label{para}

The outflow mass, moment and energy ($M_{out}$, $p_{out}$, and
$E_{out}$, respectively) were derived from line emission using the
classical formulae \citep[e.g., ][]{1985ARA&A..23..267L} as a
sum over all the velocity channels within the blue and red wings
and within the areas defined by the 3$\sigma$ contour levels of the
respective outflow lobes.  We sum all the individual pixel
contributions within the lobe area. In other words, the sum of the
contributions from each pixel and velocity channel builds up the total
outflow mass, moment and energy.  We used the $^{12}$CO(2--1),
$^{12}$CO(6--5), and $^{13}$CO(2--1) lines separately in order to have
different estimates. The abundance ratio [$^{12}$CO]/[H$_2$] and
[$^{13}$CO]/[H$_2$] was taken as $10^{-4}$ and 1.3$\times 10^{-6}$
\citep[e.g., ][]{1986ApJ...303..416S}, respectively. We conservatively assumed an
excitation temperature T$_{\rm ex}$ in the range 50-200 K for all
velocities, following the LVG analysis.  We corrected the $M_{out}$,
$p_{out}$, and $E_{out}$ for the optical depths derived in Sect. 5.1.
The average distance of 1000 pc has been used.  The $^{12}$CO molecule
is the usual outflow tracer due to its high abundance, which allows to
detect the weakest wings at the highest velocity. Nevertheless, the
risk of $^{12}$CO lies in its high opacity, which can bias the
observations towards the colder and more extended portion of the
outflow. We tried to verify this possibility by using also the
$^{12}$CO(6--5) and $^{13}$CO(2--1) lines, which present weaker
emission, thus preventing the detection of the highest velocities, but
which could possibly better trace the more collimated outflow
component.

Table \ref{energy} lists both the values derived summing the
contribution due to all the outflows present in the region (hereafter
whole emission) as well as the values of the OF1 red-shifted
outflow. In the latter case, we refer to the $^{12}$CO(2--1) line only,
since it is the only transition where the OF1 outflow is clearly
detected.  Note that, for the whole emission, we do not indicate
$\tau_{\rm blue}$ and $\tau_{\rm red}$, but, for the sake of clarity,
the mean value, $\tau_{\rm mean}$ = ($\tau_{\rm blue}$ + $\tau_{\rm
red}$)/2, which can differ slightly. However, the average values 
of the optical depth in the red- and blue-shifted lobes, $\tau_{\rm blue}$
 and $\tau_{\rm red}$,  were used in the calculations. For the $^{12}$CO(2--1) line, we adopt
$\tau_{\rm blue}=15$ and $\tau_{\rm red}=23$;  for the $^{13}$CO(2--1) transition, 
$\tau_{\rm blue}=0.2$ and $\tau_{\rm red}=0.3$; for the $^{12}$CO(6--5) line, we used a typical 
 value inferred by the LVG analysis, $\tau_{\rm blue}=\tau_{\rm red}=0.1$

From Table \ref{energy} we confirm what expected, i.e. that the
$^{12}$CO(2--1) line allows us to trace the most energetic portion of the
outflow(s) thanks to the detection of the highest velocities. The
$^{12}$CO(6--5) and $^{13}$CO(2--1) lines report $E_{\rm out}$ lower
by more than one order of magnitude. Note also that $^{13}$CO is
tracing a mass $\sim$ 2-3 times lower than that derived by the 
$^{12}$CO(2--1) transition. This is consistent with the fact that only limited velocity ranges 
of $^{13}$CO(2--1) can be used for the integrated intensity in the red-shifted
emission due to contamination from several HC tracers (see
Fig. \ref{hc}), and with the fact that the $^{13}$CO emission is more compact than that of the $^{12}$CO(2--1) line.
In addition, we cannot exclude that the mass estimate is affected by other uncertainties due
 to the presence of temperature gradients, with the
$^{13}$CO tracing warmer inner regions, as well as  to the
optical depth estimate.  Therefore, we derived the outflow kinematics
only by using the $^{12}$CO(2--1) emission.

\begin{table}
\caption{Whole emission (summing the
contribution due to the three identified outflows IRAS\,17233--OF1, --OF2 and --OF3) and  the IRAS\,17233--OF1 outflow energetics, derived assuming distance of 1000 pc 
and an excitation temperature in the 50-200 K range (see text)}
\label{energy}
\vspace{1mm}
\begin{tabular}{ccccc}
\hline
Tracer & $\tau_{m}$$^\mathrm{\dag}$ & $M_{\rm out}$$^\mathrm{\ddag}$ & $p_{\rm out}$$^\mathrm{\ddag}$ & $E_{\rm out}$$^\mathrm{\ddag}$ \\
       &                            & ($M_{\odot}$) & ($M_{\odot}$~km s$^{-1}$) & ($10^{45}$ erg) \\
\hline
\multicolumn{5}{c}{Whole emission (OF1+OF2+OF3)} \\
\hline
$^{12}$CO(2--1) & 19.0 & 2.4-3.7 & 99-140 & 60-90 \\
$^{13}$CO(2--1) &  0.3 & 1.0-1.5 & 19-30  & 4-6  \\
$^{12}$CO(6--5) &  0.1 & 0.7-1.1 & 20-29  & 6-10  \\
\hline
\multicolumn{5}{c}{OF1 red-shifted outflow} \\
\hline
$^{12}$CO(2--1) & 23.0 & 0.6-0.9 & 18-28 & 6-9 \\
\hline
\end{tabular}
\begin{list}{}{}
\item[$^\mathrm{\dag}$] \footnotesize{Whole emission: mean $\tau$ between blue and red emission optical depths (Sect. 5.1)}
\item[$^\mathrm{\ddag}$] \footnotesize{Values derived  for T$_{\rm ex}$  50 and 200 K (Sect. 5.2)}
\end{list}
\end{table}

In order to derive the real outflow parameters, the angle $\theta$ of
inclination to the plane of the sky has to be assumed. Unfortunately,
because of the complexity of the region and the outflow multiplicity,
we cannot derive a good estimate of $\theta$. However, taking into
account the quite elongated structure of the outflow(s) and the very
limited overlap between red and blue lobes, it is reasonable to assume
an intermediate inclination, in the 30$^\circ$--60$^\circ$
range.  A kinematic age for each flow lobe, $t_{\rm kin}$, was
determined from the projected distance between the position of the
outflow lobe and that of the mm-peak, and the corresponding radial
outflow velocity component.  From the kinematic outflow age, the mass
entrainment rate of the molecular outflow, $\dot{M}_{out}$, the
momentum rate, $\dot{p}_{tot}$, and the mechanical luminosity,
L$_{mec}$, were derived for the blue and red lobes separately, and
then added together to obtain the total values presented in Table
\ref{energy2}. There, we
list  $t_{\rm kin}$, $\dot{M}_{out}$, $\dot{p}_{tot}$, L$_{mec}$
without correction for the inclination angle, but also for $\theta$
equals 30$^\circ$ and 60$^\circ$.  As done for optical depths, we
give a mean value for the kinematic age between the values derived
separately for the blue and red emission. After geometry correction, for the
whole emission we have a kinematic age of 10$^2$--10$^3$ yr, and
consequently $\dot{M}_{out}$ $\simeq$ 2--9 10$^{-3}$ $M_{\odot}$
yr$^{-1}$, $\dot{p}_{\rm tot}$ $\simeq$ 0.1--0.7 $M_{\odot}$~km
s$^{-1}$ yr$^{-1}$, and $L_{\rm mec}$ $\simeq$ 500--8000 $L_{\odot}$.
By comparing these estimates with the expected correlation between the
dynamical parameters and the source bolometric luminosity
\citep[e.g.,][]{2002A&A...383..892B}, and taking into account the
outflow multiplicity, we infer for the driving sources a $L_{\rm bol}$
$\ge$ 10$^{4}$ $L_{\odot}$, i.e. earlier than a B0.5 type.  On the
other hand, if we take into account the OF1 outflow only, the bolometric luminosity of the driving source
 is  of the order of 10$^{4}$ $L_{\odot}$.  The luminosity
estimated for the driving sources are well in agreement with the
bolometric luminosity of IRAS\,17233, which is 2.5 $\times 10^4$~L$_\odot$ for a
distance of 1 kpc.

The derived values are in general good agreement with those 
($M_{\rm out}=0.8~M_{\odot}$, $p_{\rm out}=25.3$~$M_{\odot}$~km s$^{-1}$, $E_{\rm out}=8.6\times 10^{45}$~erg, 
$t_{\rm kin}=6200$~yr)
reported
by \citet{2008A&A...485..167L} based on the analysis of single dish data
in CO(3--2), and with the conclusion that the kinematical outflow
parameters are typical of massive YSOs. The differences between the
two analyses can be explained  with the different areas 
and different velocities adopted for
the analysis. Moreover, our previous study on CO(3--2) did not include
any correction for the opacity of the line.

\subsection{Age vs. outflow collimation}

The collimation and the age of the outflows can be compared with the
scenario described by \citet{2005ccsf.conf..105B} where jet-like
outflows occur only in early evolutionary phases where no HCH{\sc ii} region has formed and with stellar luminosities 
 corresponding to late B-type main sequence values, whereas
in later evolutionary phases, a HCH{\sc ii} forms and the wind from
the central massive star produces an additional less collimated
outflow component.  Since to form early O-type massive stars
via accretion the protostellar objects must accrete even after the central object
has reached the main sequence, the evolutionary scenario depicted above also corresponds 
to a change of luminosity of the (proto)star during its evolution, 
from a late B-type to an early O-type star  before reaching its final mass and stellar luminosity. 
This scenario would explain the observational result that no collimated jet-like outflow has ever been detected 
from very young early O-type (proto)stars.

In the IRAS\,17233 case, the N-S outflows show a remarkable degree of
collimation ($f_c \sim 4$) as well as a relatively small opening
angle, $\delta \sim 11\degr$.  In addition, these estimates have to be
considered as lower (collimation) and upper (opening angle) limits,
given the unknown outflow multiplicity.  The spectral type
(B0.5) derived by the outflow parameters and the age of flow ($\sim$ 10$^2$-10$^3$ yr) 
agree with the scenario of  \citet{2005ccsf.conf..105B}. 
However, the OH maser spots
trace a bipolar outflow with a definitely larger opening angle (see
Fig.~\ref{con}) than that observed in CO and H$_2$, thus suggesting we
are observing a stage where jet-like and less collimated winds
co-exist.  The detection of  HCH{\sc ii}, typical of more evolved objects, 
also supports our interpretation that IRAS\,17233 is in a transitional phase, when the star also produces an additional
less collimated outflow component.

\begin{table}
\caption{Whole emission (due to the three identified outflows IRAS\,17233--OF1, --OF2 and --OF3) and IRAS\,17233--OF1 outflow timescales and kinematics, derived from $^{12}$CO(2--1) emission, 
assuming distance of 1000 pc (see text). Uncorrected values as well as values corrected for
assumed inclinations to the plane of the sky ($\theta$) are reported}
\label{energy2}
\vspace{1mm}
\begin{tabular}{ccccc}
\hline
$\theta$ & $t_{\rm kin}$$^\mathrm{\dag}$ & $\dot{M}_{out}$$^\mathrm{\ddag}$ &  $\dot{p}_{\rm tot}$$^\mathrm{\ddag}$ & $L_{\rm mec}$$^\mathrm{\ddag}$ \\
 ($\deg$) & (yr) & ($M_{\odot}$ yr$^{-1}$) & ($M_{\odot}$~km s$^{-1}$  yr$^{-1}$) & ($L_{\odot}$) \\ 
\hline
\multicolumn{5}{c}{Whole emission (OF1+OF2+OF3) } \\
\hline
 -- & 480-940 & 3-5 10$^{-3}$ & 0.1-0.2 & 760-1200  \\
 30 & 278-545 & 5-9 10$^{-3}$ & 0.4-0.7 & 5244-8280  \\
 60 & 830-1626 & 2-3 10$^{-3}$ & 0.09-0.13 & 585-924  \\
\hline
\multicolumn{5}{c}{OF1 red-shifted outflow} \\
\hline
 -- & 360-600 & 1-2 10$^{-3}$ & 0.03-0.05 & 86-132  \\
 30 & 209-348 & 2-3 10$^{-3}$ & 0.1-0.2 & 593-911  \\
 60 & 623-1038 & $\sim$ 1 10$^{-3}$ & 0.02-0.03 & 66-102 \\
\hline
\end{tabular}
\begin{list}{}{}
\item[$^\mathrm{\dag}$] \footnotesize{Whole emission: mean $t_{\rm kin}$ between blue and red emission (Sect. 5.1)}
\item[$^\mathrm{\ddag}$] \footnotesize{Values derived for T$_{\rm ex}$  50 and 200 K (Sect. 5.2)}
\end{list}
\end{table}

\section{Conclusions}

We presented  interferometric (SMA) and single-dish (APEX) 
observations of one of the nearest massive star forming region
IRAS\,17233 in CO(2--1) and (6--5) with angular
resolution of $5\farcs4 \times 1\farcs9$ and $9\farcs4$, respectively.
The main results can be summarised as follows:

\begin{enumerate}

\item
The data reveal a clumpy extended ($\sim$ 50$\arcsec$) structure with
well separated blue- and red-shifted emission, and an overall
structure roughly aligned along the N-S direction and centred on the
region where groups of H$_2$O, CH$_3$OH, and OH maser spots are
detected.  This region is also associated with the peak of the 1.3~mm
emission we mapped, as well as, on subarcsec-scale, with 4 HCH{\sc ii}
regions (called VLA~2a, b, c and d) observed at cm-wavelengths and
indicating massive star formation.  The outflow is associated with
extremely high velocities, up to $\sim$ --200 and +120 km s$^{-1}$
with respect to the ambient LSR velocity.  Up to eight outflow clumps have been
observed in CO(2--1).  The whole outflow structure is very well collimated ($f_c
\sim 4$) and with an opening angle of $\sim 11\degr$.

\item
The large number of outflow clumps clearly indicates  outflow multiplicity, as expected
for a cluster of high-mass YSOs.
Using high-angular resolution maps of H$_2$, cm-continuum, and maser emission,
we tried to identify different outflow components. At least three outflows can
be individuated. One outflow (IRAS\,17233-OF1) is compact ($\sim$ 5$\arcsec$-10$\arcsec$) and located
along the NW-SE direction: it shows extremely 
high-velocity red-shifted emission and it is associated with a jet traced by H$_2$ emission
as well as with red and blue H$_2$O maser spots and SO emission.
Its driving source has to be found in the region associated with the 1.3~mm object,  
the CH$_3$OH masers, and two massive YSOs (VLA~2a and VLA~2b).  
On the other hand, the extended ($\ge$ 50$\arcsec$; $\ge$ 0.2 pc) emission  along the N-S direction can be
disentangled into at least two more bipolar outflows: one (IRAS\,17233-OF2) associated 
with a counterpart on smaller scales ($\sim5''$; $\sim$5000~AU) traced by red- and blue-shifted OH masers
and driven by a massive YSO traced by cm-emission (VLA~2c or VLA~2d), 
and another one (IRAS\,17233-OF3) whose driving source has to be found again among VLA~2a and the 1.3~mm peak.

\item
H$_2$ emission is found mostly in correspondence to the blue-shifted clumps,
probably due to extinction hiding the red-shifted H$_2$ lobes. In two cases, we found extremely high velocity
CO emission associated with H$_2$ emission, suggesting we are observing
material accelerated along the jet traced by H$_2$.

\item
The molecular outflows were observed also in
 isotopologues of $^{12}$CO: i.e. $^{13}$CO(2--1) and C$^{18}$O(2--1), as well as in
SO(6$_{\rm 5}$-5$_{\rm 4}$). Although covering a smaller velocity range with respect to
the $^{12}$CO(2--1) line (due to lower line brightness), also in this case
the outflow structure is well traced.
By comparing the intensity of the $^{13}$CO and $^{12}$CO(2--1) wings,
we have estimated the optical depths: the emission in
the $^{13}$CO(2--1) line results optically thin, with
$\tau_{\rm 13}$ $\le$ 0.1-0.3. Consequently, the $^{12}$CO(2--1) optical depth 
(derived simply scaling by the assumed [$^{12}$CO/$^{13}$CO] ratio) is quite
high ($\sim$10-30), indicating thick emission even at outflow velocities $\pm$ 30-60
km s$^{-1}$ with respect to the ambient emission. 

\item
To estimate the relative $^{12}$CO excitation conditions of the outflow clumps, we compared
the $J$~=~6--5 and 2--1 line brightnesses. For the blue-shifted
lobe the higher  the distance of the clump from the driving source(s), the
higher is the $R_{\rm 62}$ ratio, suggesting higher excitation conditions.
In addition, at least for the red clumps, we find  indication of an increase
of the excitation at the highest velocities, probably closely associated
with recently shocked gas.

\item
Although only with two transitions,
LVG statistical-equilibrium calculations  give a rough
estimate of the physical parameters of the molecular gas traced by CO.  The red and blue clumps closer to the YSOs are
associated with $T_{\rm kin}$ $\ge$ 50 K and with a gas pressure
$\ge$ 10$^6$ K cm$^{-3}$. 

\item
The estimate of the kinematical outflow parameters such as momentum,
kinetic energy, and the mechanical luminosity show values that, even
taking into account the outflow multiplicity, are typical of massive
YSOs ($L_{\rm bol}\ge 10^4$~L$_\odot$), and in agreement with the
measured whole bolometric luminosity of the source ($L_{\rm
bol}=2.5~10^4$~L$_\odot$).

\item
The kinematic ages of the flows are in the range $10^2-10^3$ yr, and
therefore point to young objects that still did not reach the main
sequence. We compared our results with the scenario described by
\citet{2005ccsf.conf..105B} where jet-like outflows are powered only
by young protostars with luminosities that correspond to late B-type main sequence values,  and still without an HCH{\sc ii}
region. In the case of the N-S outflows in IRAS\,17233, we observe a
high degree of collimation as well as the indication of definitely
less collimated wind traced by OH maser spots.  We thus suggest that we
are observing a stage where jet-like and less collimated wind
co-exist, in agreement with the detection of very small ($\sim$
10$^{-3}$ pc) ionised regions and by the spectral type (B0.5) derived
from the outflow parameters.

\item
Our analysis shows the importance of high sensitivity observations to
detect extremely high velocity emission to access  the most energetic
part of molecular outflows and detect the counterpart of the H$_2$ emission.
Moreover, high-spatial resolution observations of higher excitation CO lines
as well as SiO,  sensitive to the hot jet component and not contaminated by ambient gas emission as 
low rotational CO transitions, would be instructive to better assess the outflow multiplicity in the source and for a better
characterisation of the excitation conditions.

\end{enumerate}

\begin{acknowledgements}
The Submillimeter Array is a joint project between the Smithsonian
Astrophysical Observatory and the Academia Sinica Institute of
Astronomy and Astrophysics and is funded by the Smithsonian
Institution and the Academia Sinica.
\end{acknowledgements}

\end{document}